\documentclass{jaa}
\pdfoutput=1
\usepackage{natbib}
\usepackage{aas_macros}

\usepackage{bm}
\usepackage{graphicx}
\usepackage{siunitx}
\usepackage{hyperref} 
\hypersetup{colorlinks=true,linkcolor=black,anchorcolor=black,citecolor=black,filecolor=black,menucolor=black,runcolor=black,urlcolor=black}
\usepackage{amsmath}
\usepackage[bottom]{footmisc}
\usepackage{chngcntr}

\begin{document}\sloppy

\title{The role of magnetic fields in the fragmentation of the Taurus B213 filament into Sun-type star-forming cores}

\author{R. Anirudh,\textsuperscript{1,*} Chakali Eswaraiah,\textsuperscript{1,*}\footnote{Ramanujan Fellow}~~~Sihan Jiao,\textsuperscript{2,3}\and Jessy Jose\textsuperscript{1}}
\affilOne{\textsuperscript{1}Indian Institute of Science Education and Research (IISER) Tirupati, Rami Reddy Nagar, Karakambadi Road, Mangalam (P.O.), Tirupati 517 507, India \\} 
\affilTwo{\textsuperscript{2}CAS Key Laboratory of FAST, National Astronomical Observatories, Chinese Academy of Sciences, People's Republic of China \\} \affilThree{\textsuperscript{3}University of Chinese Academy of Sciences, Beĳing 100049, People’s Republic of China \\}

\twocolumn[{

\maketitle

\corres{ranirudh@students.iisertirupati.ac.in, eswaraiahc@labs.iisertirupati.ac.in}

\msinfo{}{}

\begin{abstract}
Fragmentation is a key step in the process of transforming clouds (and their substructures such as filaments, clumps, and cores) into protostars. The thermal gas pressure and gravitational collapse are believed to be the primary agents governing this process, referred to as the thermal Jeans fragmentation. However, the contributions of other factors (such as magnetic fields and turbulence) to the fragmentation process remain less explored. In this work, we have tested possible fragmentation mechanisms by estimating the mean core mass and mean inter-core separation of the B213 filament. We have used the $\sim\ang{;;14}$ resolution James Clerk Maxwell Telescope (JCMT) Submillimetre Common-User Bolometer Array 2 (SCUBA-2)/POL-2 850 $\mu$m dust continuum map and combined it with a {\it Planck} 850 $\mu$m map and {\it Herschel} data. We find that in addition to the 
thermal contribution, the presence of ordered magnetic fields could be important in the fragmentation of the B213 filament.
\end{abstract}

}]


\doinum{12.3456/s78910-011-012-3}
\artcitid{\#\#\#\#}
\volnum{000}
\year{0000}
\pgrange{1--10}
\setcounter{page}{1}
\lp{10}

\section{Introduction}

Observations have shown that filamentary structures are ubiquitous in molecular clouds \citep{Andre2014,Tafalla2015}. These filaments fragment to form dense cores, which further collapse to form protostars. This fragmentation is modeled using gravitational instability, turbulence, and magnetic fields. The Jeans instability relations \citep{Jeans} govern the contributions of the thermal gas pressure and the gravitational potential. These relations determine the critical length scales\footnote{Critical lengths refer to threshold characteristic scales analogous to the Jeans length beyond which a self-gravitating cloud is unstable to perturbations. This nomenclature is widely used instead of the Jeans length for models incorporating magnetic fields/compressibility of the fluid.} and masses at which gravitational collapse occurs. However, they are only accurate for clouds without turbulence and magnetic fields. The magneto-hydrostatic equilibrium and stability of a gas cloud generally depend on the amplitude of external perturbations acting on it.\\

When magnetic field contributions are considered, the effects depend on its orientation and strength as well as on the geometry of the cloud. Dynamical instability occurs when the magnetic energy (due to a uniform, isotropic magnetic field) exceeds the gravitational potential energy. For an axial magnetic field\footnote{Axial magnetic fields refer to fields parallel to a cylinder's long axis (in this case, a gas filament).} through an infinitely long, incompressible cylinder\footnote{The condition of incompressibility is $\nabla . V$=0, where $V$ is the fluid's velocity field (gas). Compressible fluids refer to systems that do not follow this relation while solving for the Navier-Stokes equations.}, the system is unstable for transverse oscillations with a wavelength of perturbation exceeding a critical value \citep{ChandrasekharFermi}. The magnetic field does not affect the equilibrium of the fluid but only acts on the perturbation imposed. At equilibrium, the gravitational instability is counteracted by gas (sum of thermal and non-thermal contributions) and magnetic pressures. Since the latter increases with field strength, the critical length scale (of perturbation) increases with increasing magnetic field strength.\\


However, this only holds if the system is incompressible. For a compressible cylinder, the transversal (in the direction parallel to the filament minor axis) dimensions expand with increasing field intensity. This expansion leads to a rise in excess gravitational pressure, thereby causing instability in the system. This indirect contribution to gravitational instability is stronger than the 
stability rendered by magnetic pressure. 
Thus, for a compressible cylinder with an axial magnetic field, the critical length of perturbation decreases with increasing magnetic field strength \citep{Stodolkiewicz}. For zero magnetic field intensity, the critical wavelength ($\mathrm{\lambda_{crit}}$) is 1.77 times the Jeans length \citep{Stodolkiewicz}.\\ 


In contrast, \citet{Hanawa2017} demonstrated that magnetic fields work against fragmentation for an isothermal cylinder permeated by a perpendicular magnetic field that is affected by a sinusoidal perturbation. They also show that magnetic fields reduce the growth rate of a perturbation for the free boundary condition but cannot fully stabilize the system as the growth rate remains positive even for an infinitely strong magnetic field. Thus, the critical wavelength and length scale of fragmentation increases above the zero field intensity $\mathrm{\lambda_{crit}}$ of 1.77$\mathrm{\lambda_{J}}$ indicated by \citet{Stodolkiewicz}. This length scale depends on the magnetic field strength and the perturbation.\\

Another factor relevant to filament fragmentation is the presence of turbulence \citep[see][]{Federrath&Klessen2012,Jappsen2005,Gehman1996}. Turbulence increases the growth rate and length scale of the fastest growing mode for weak magnetic fields. Strong magnetic fields produce an opposite effect, and decrease the growth rate of the fastest growing mode \citep{Gehman1996}. Above a threshold, the magnetic field reduces the length scale of fragmentation in turbulent clouds.\\

In more recent theoretical work, \citet{Clarke2017} used smoothed particle hydrodynamical (SPH) simulations to model filaments fragmenting due to gravitational collapse and accretion from the turbulent medium. They reported that increased turbulence dominated gravitational effects, changed fragmentation length scales, and caused elongated fiber-like filaments to form within the larger filament - which is similar to the {\it fray and fragment} scenario proposed by \citep{Tafalla2015}. They also concluded that large gradients in the vorticity of the velocity field caused more gas to accrete to form dense structures.\\

This article focuses on testing the fragmentation mechanism occurring in the L1495/B213 filament of the Taurus molecular cloud, a widely studied star-forming region. At a distance of about 140 pc \citep{Elias1978}, it is one of the closest Sun-type star-forming regions. It is an excellent candidate for studying the fragmentation scenario using the sensitive and high-resolution dust continuum maps acquired from the JCMT SCUBA-2/POL-2 BISTRO Survey \citep{Eswaraiah}.\\ 

The description of the data we utilized for this work is given in Section \ref{sec:data} The results based on the identification of cores are shown in Section \ref{sec:results}  The implications of the core properties and fragmentation analysis are discussed in Section \ref{sec:Discussion} The article is summarised in Section \ref{sec:Summary}

\begin{figure*}[t]
    \centering
    \includegraphics[width=14 cm, height=14 cm]{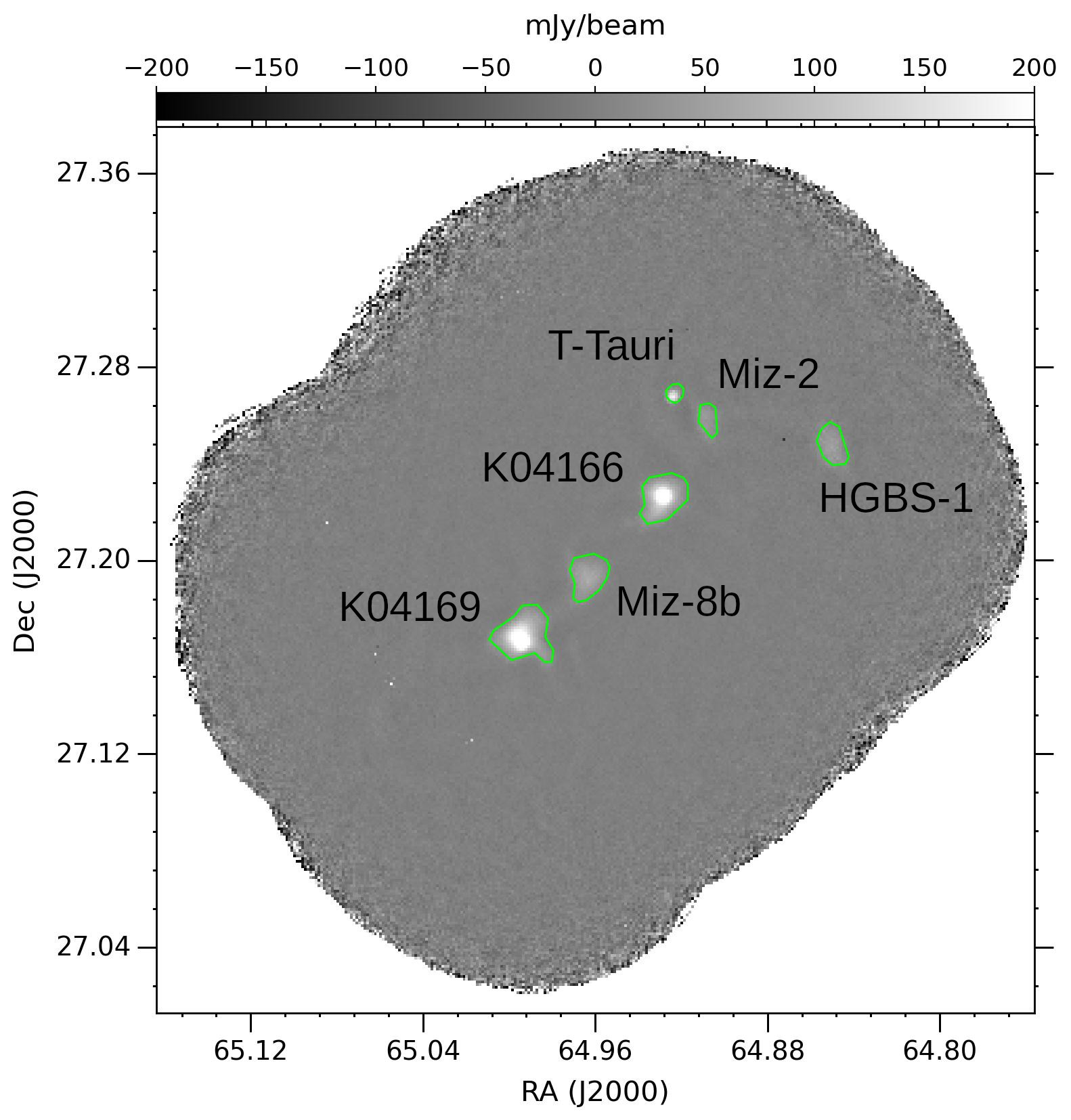}
    \caption{JCMT SCUBA-2/POL-2 dust continuum map of the Taurus B213 filament at 850 $\mu$m.  The five cores and the T-Tauri star are labeled. The green contours represent the core boundaries obtained based on the FellWalker algorithm.} 
    \label{fig:polygon}
\end{figure*}

\vspace{-1 em}

\begin{table*}[t]
\centering
\caption{Spatial properties of the ellipse fits to the cores obtained from the JCMT map}
\label{tab:coreparameters}
\begin{tabular}{|p{7cm}|c|c|c|c|c|}
\hline
\textbf{Core properties} & \textbf{K04166} & \textbf{K04169} & \textbf{Miz-8b} & \textbf{HGBS-1} & \textbf{Miz-2} \\ \hline
Centre of the core RA (deg)           & 64.928      & 64.994        & 64.965       & 64.851      & 64.906     \\ \hline
Centre of the core Dec (deg)          & 27.226      & 27.168        & 27.192        & 27.250      & 27.255     \\ \hline
Position Angle (N to E; deg)       & 127          & 126          & 119         & 20      & 17       \\ \hline
Semi-major axis (a) (pc)           & 0.029       & 0.032       & 0.029      & 0.022      & 0.019     \\ \hline
Semi-minor axis (b) (pc)           & 0.0207      & 0.021      & 0.022      & 0.013      & 0.01     \\ \hline
Offset angle between the filament$\mathrm{^a}$ and the cores (deg) & 6 & 7 & 14 & 113 & 116  \\ \hline
Offset angle between the large-scale mean B-field$\mathrm{^b}$ and the cores (deg) & 98 & 97   & 90 & 9 & 12  \\ \hline
Dust temperature of the core (in K) & 12.2$\pm$0.4	& 12.8$\pm$0.5 & 10.7$\pm$0.1 & 11.8$\pm$0.1 & 11.4$\pm$0.1     \\ \hline
\end{tabular}\\
Notes:\\
$\mathrm{^a}$ Filament position angle is $\ang{133}$\\
$\mathrm{^b}$ Large-scale B-field orientation $\ang{29}$
\end{table*}

\section{Data}\label{sec:data}

The B213 filament consists of a chain of two protostellar cores (K04166 and K04169), three prestellar cores (Miz-8b, Miz-2, and HGBS-1), and a T-Tauri star 
as shown in Figure \ref{fig:polygon}. It is shifted away from the straight L1495 filament and exhibits a kink-like structure \citep[see Figure 1(a) of][]{Eswaraiah}.  The position angle (N to E) of the filament is $\ang{133}$, and the filament takes a $\ang{90}$ turn towards the core HGBS-1 as shown in Figure 3 of \citet{Palmeirim2013}. The mean large-scale magnetic field orientation is $\ang{29}$ \citep{Eswaraiah}.\\


For this work, we have used an 850 $\mu$m thermal dust emission map acquired by the JCMT telescope equipped with SCUBA2 camera and POL2 polarimeter. The polarization data was acquired as a part of the B-fields In STar-forming Region Observations (BISTRO; \citet{Ward-Thompson2017}; JCMT project code M16AL004) survey between November 5th 2017 and January 8th 2019. The two fields were observed (each of them 20 times) using the POL-2 DAISY mapping mode \citep{Fribergetal2016}. This mode results in maps with a $\ang{;12;}$ diameter, of which the central $\sim$7 represents usable coverage, so these two pointings represent a tightly spaced mosaic. These
observations were made in JCMT weather bands 1 and 2, with the 225 GHz atmospheric opacity ($\tau_{225}$) varying between 0.02 and 0.06. The total exposure time for the two fields is $\sim$28 hr (14 hr in each of the two overlapping fields), resulting in one of the deepest observations yet made by the BISTRO survey.\\

The 850 $\mu$m POL-2 datasets were reduced using the pol2map routine recently added to SMURF \citep{Berryetal2005,Chapinetal2013}. The final mosaicked maps, calibrated in milliJanskys per beam, are produced from coadded Stokes I, Q, and U maps
with a pixel size of $\ang{;;4}$. The final rms noise for the Stokes I map binned to a pixel size of $\ang{;;12}$ was $\sim$1.3 mJy/beam.
We used the Stokes I map for this work, i.e., the thermal dust continuum map shown in Figure \ref{fig:polygon}. 
More details on the POL2 data reduction and analyses can be found in one of the recent BISTRO papers \citep{Chingetal2022}.\\



Given the close proximity of the Taurus molecular cloud, the observations resolve the B213 filament into individual cores at a resolution of $\ang{;;14.1}\simeq$0.01 pc at 140 pc.\\ 

The work presented here also uses data from other facilities to extract the filament properties to perform Jeans fragmentation analyses.
We use the dust temperature map and the column density map ($\ang{;;36.3}$ and $\ang{;;18.2}$ respectively) obtained by \citet{Palmeirim2013} as part of the Herschel Gould Belt Survey \citep{HGBS, Marsh2016}\footnote{\url{http://www.herschel.fr/cea/gouldbelt/en/Phocea/Vie_des_labos/Ast/ast_visu.php?id_ast=66}}. These maps are made using the Herschel PACS 160 $\mu$m band and the SPIRE 250, 350, and 500 $\mu$m bands. This data is then combined with a Planck 353 GHz map using the J-comb algorithm \citep{Jiao2022} to obtain an improved column density map, thereby recovering extended low-density material.\\

The J-comb algorithm combines high and low-resolution images linearly and aids in the easier convolution of multi-wavelength data products. It achieves this by applying a taper function\footnote{A taper function is a non-constant scaling function that modifies and combines low and high-resolution beams. Mathematically, it is the product of the amplitude modulation factor and a weighting function. This method provides a meaningful approach to combining the contributions of low-density and high-density materials. See equations 17 and 18 of \citep{Jiao2022}} to the low-pass filtered image and combine it with weighted high-pass images. Unlike algorithms that do not preserve the beam shapes of the original maps, this method guarantees a near-Gaussian beam shape for the combined map. \citet{Jiao2022} demonstrated that this algorithm performs better than CASA-feather and Miriad-IMMERGE using benchmark tests and also asserted that J-comb shows greater detail than earlier studies for combining data products from ground-based and space-based observatories.

\section{Results}\label{sec:results}

\subsection{Identification of cores}

The five cores (K04166, K04169, Miz-8b, HGBS-1, Miz-2) can be distinguished from the rest of the filament due to their higher density and emission. The FellWalker algorithm  \citep{Berry2015} of the CUPID package (part of the Starlink project) is used to identify clumps\footnote{In this context, ``clumps" do not refer to cloud substructures that are larger than 0.1 pc. They refer to ``clumps of emission" corresponding to star-forming cores in the filament.} of emission and obtain their dimensions. The algorithm takes several input parameters, such as root mean square (RMS) noise, beam size (FWHM), the threshold value for clump peak, maximum corrupted pixels allowed inside a clump, etc., to identify the cores using the steepest ascent (gradient-tracing) approach. The threshold clump peak value is generally taken as twice the RMS noise. Using the 850 $\mu$m POL-2 map and by setting the parameters for the FellWalker algorithm, the star-forming cores are identified as contours on the POL-2 map.\\

The output of the FellWalker algorithm is a set of 5 parameters describing the properties of the cores -- (i) RA and Dec values of vertices of the core polygons, (ii) centers, (iii) areas, (iv)  spatial extents in RA and Dec, and (v) peak dust emission values. 
These polygons, which physically trace the core boundary, are fitted as an ellipse using the SciPy least squares method. The best-fit yields the dimensions of the semi-major and semi-minor axes, the center, and the position angle of a given core, and are listed in Table \ref{tab:coreparameters}.
The identified cores and polygons obtained from the FellWalker algorithm are shown in Figure \ref{fig:polygon}. \citet{Berry2015}\footnote{and also Watson 2010 - http://hdl.handle.net/2299/13228} conclude that for 2-D artificial Gaussian clumps of emission, the FellWalker algorithm is less dependent on specific parameter settings and less likely to create false clump detections or split up large clumps as compared to CLUMPFIND. This motivated us to use the FellWalker algorithm over CLUMPFIND for identifying star-forming cores.\\

It is important to note that the core K04169 does not resemble the shape of an ellipse. It has a protrusion towards the South-West direction as seen in Figure \ref{fig:polygon}. However, the masses calculated from the integrated column densities for the ellipse and the polygon are within the error limits in question. Therefore, fitting an ellipse for K04169 is still valid. The other four cores resemble the shape of an ellipse due to their prolate-like or oblate-like structures. Hence, an elliptical fit is a meaningful choice.

\subsection{Core properties}

Using the fitted ellipses (Figure \ref{fig:ellipses}), it is possible to estimate the mass of the cores based on the Herschel column density map and POL-2 dust continuum map. Equation \ref{eq:columnmass} is used to calculate the mass of the cores from the column density map, where M is the mass of the core in kg, $\mathrm{\mu}$ is the mean molecular weight considered as 2.8 based on cosmic abundances \citep{Kaufmann2008}, $\mathrm{m_{H}}$ is the mass of the Hydrogen atom, A is the area of a pixel and $\mathrm{N(H_{2})}$ is the column density of a pixel. The area of each pixel is calculated by converting the RA/Dec pixel area of 3$\times$3 arcsec$^{2}$ to cm$^{2}$ units. This is done by using the angular separation formula shown in equation \ref{eq:angularsep} (assuming a constant distance to the filament of 140 pc). From this calculation, A is equal to $\mathrm{3.948\times 10^{31}\,cm^2}$ on the projected sky at a distance of B213.\\

Therefore, the integrated column density is the summation of the masses corresponding to every pixel within the fitted ellipse of the core. Since the mass of each pixel is the product of the constant parameters $\mathrm{\mu}$, $\mathrm{m_{H}}$ and $\mathrm{A}$, and the variable $\mathrm{N(H_{2})}$,   
we numerically add the column density values of all the pixels within a core and label that as $\mathrm{N(H_{2})_{sum}}$. Thus,
\begin{equation}
\begin{split}
    M = \sum_{i}\left(\mu\,m_{H}\,A\,N(H_2)_i\right) \\= \mu\,m_{H}\,A\,\sum_{i}N(H_2)_i \\= \mu\,m_{H}\,A\,N(H_{2})_{sum}, 
    \label{eq:columnmass}
\end{split}
\end{equation}
    
where $i$ is the summation index for the pixels. \\

The masses of the cores can also be calculated from the POL2 dust continuum map using the following equation \citep{Hildebrand1983}
\begin{equation}
    M = \frac{F_{\nu}D^{2}}{B_{\nu}(T_{d})\kappa_{\nu}},
    \label{eq:fluxmass}
\end{equation}
where $\mathrm{F_{\nu}}$ is the integrated flux in the fitted ellipse of the core. $\mathrm{T_{d}}$ is the mean dust temperature obtained by computing the arithmetic mean of the dust temperature of all pixels within the respective core's boundary. D is the distance to the B213 filament taken as 140 pc, $\mathrm{B_{\nu}}$ is the Planck function value at the specified dust temperature and observed frequency, $\mathrm{\kappa_{\nu}}$ is the dust mass opacity, taken to be equal to 0.00125 $\mathrm{m^2\,kg^{-1}}$  \citep{Johnstone2017}, and the frequency $\mathrm{\nu}$ corresponds to a wavelength of $\mathrm{850\,\mu m}$. Table \ref{tab:coreparameters} shows the core properties, dust temperatures, etc. Table \ref{tab:columnmasstable} contains the core masses obtained from the Herschel column density map. Table \ref{tab:fluxmasstable} contains the core masses obtained from the POL-2 map.

\begin{table}[h]
\centering
\caption{Masses of the cores obtained from the Herschel map}
\label{tab:columnmasstable}
\begin{tabular}{|c|c|c|}
\hline
\textbf{Core} & $\bm{\mathrm{N(H_2)_{sum}}}$ $\bm{(\times \mathrm{10^{24}, cm^{-2}})}$ & \textbf{Mass $(\bm{\mathrm{M_{\odot}})}$} \\ \hline
K04166                    & 6.5                                                          & 0.6                                   \\ \hline
K04169                    & 8.5                                                          & 0.8                                   \\ \hline
Miz-8b                    & 5.0                                                          & 0.4                                   \\ \hline
HGBS-1                    & 3.5                                                          & 0.3                                   \\ \hline
Miz-2                     & 1.7                                                          & 0.2                                   \\ \hline
\end{tabular}
\end{table}

\vspace{-1 em}

\begin{table}[h]
\centering
\caption{Masses of the cores obtained 
from the JCMT SCUBA-2/POL-2 dust continuum map}
\label{tab:fluxmasstable}
\begin{tabular}{|c|c|c|}
\hline
\textbf{Core} & \textbf{Integrated Flux (mJy)} & \textbf{Mass $\mathrm{M_{\odot}}$)} \\ \hline
K04166                    & 1502                        & 0.5$\pm$0.3                                   \\ \hline
K04169                    & 1951                        & 0.6$\pm$0.3                                   \\ \hline
Miz-8b                    & 484                         & 0.2$\pm$0.1                                   \\ \hline
HGBS-1                    & 291                         & 0.10$\pm$0.05                                   \\ \hline
Miz-2                     & 137                         & 0.05$\pm$0.03            \\ \hline
\end{tabular}
\end{table}

There is a clear discrepancy between the masses obtained from the Herschel column density map and the JCMT POL-2 dust continuum map. This inconsistency can be attributed to the fact that there is a missing flux in the POL-2 map, possibly due to spatial filtering of the extended emission. The masses obtained from the POL-2 map are lesser than those obtained from the Herschel column density map. This further supports the hypothesis.\\

To solve this flux issue, maps from three telescopes (JCMT, {\it Herschel}, and {\it Planck}) are merged using the J-comb algorithm to obtain a corrected column density map \citep[see][]{Jiao2022}. The FellWalker algorithm is then run on this map to estimate the spatial dimensions of the cores (Figure \ref{fig:combined}). The masses of the identified cores are then estimated using equation \ref{eq:fluxmass} and are shown in Table \ref{tab:correctedmasstable}. We use these corrected masses from the J-comb combined map for comparison with the Jeans mass.\\

The uncertainties in mass for the POL-2 map and the combined map are estimated by propagating the standard deviation in $\mathrm{T_{d}}$\footnote{computed by finding the standard deviation of the dust temperature values of all pixels inside the respective core, shown in Table \ref{tab:coreparameters}}, 10$\%$ of the $\mathrm{F_{\nu}}$ value as the flux calibration
uncertainty of SCUBA-2 \citep{Dempsey2013}, and a 50$\%$
uncertainty in dust mass opacity \citep{Roy2014}. These uncertainties are also listed in Tables \ref{tab:fluxmasstable} and \ref{tab:correctedmasstable}.

\vspace{-1em}

\begin{table}[h]
\centering
\caption{Masses of the cores obtained from the J-comb combined map}
\label{tab:correctedmasstable}
\begin{tabular}{|c|c|c|}
\hline
\textbf{Core} & \textbf{Integrated Flux (mJy)} & \textbf{Mass ($\mathrm{M_{\odot}}$)} \\ \hline
K04166                    & 2223                        & 0.8$\pm$0.4                                   \\ \hline
K04169                    & 3180                        & 1.0$\pm$0.5                                   \\ \hline
Miz-8b                    & 1144                         & 0.5$\pm$0.3                                   \\ \hline
HGBS-1                    & 2918                         & 1.1$\pm$0.5                                   \\ \hline
Miz-2                     & 2437                         & 0.9$\pm$0.5                                   \\ \hline
\end{tabular}
\end{table}

\vspace{-1em}
\subsection{Filament parameters and fragmentation analyses}

Using the RA and Dec values of the centers of the cores and setting the distance to the B213 filament as 140 pc, it is possible to estimate the inter-core separations using the following formulae:

\begin{subequations}
    \begin{equation}
        \theta =\cos ^{-1}\left[\sin \delta _{A}\sin \delta _{B}+
    \cos \delta _{A}\cos \delta _{B}\cos(\alpha _{A}-\alpha _{B})\right]
    \end{equation}
and

    \begin{equation}
        c^2=a^2+b^2-2ab\,\cos(\theta),
    \end{equation}
    \label{eq:angularsep}
\end{subequations}

where $\theta$ is the angular separation of two points on the surface of a sphere (celestial sphere defined using the equatorial coordinate system). The $\alpha_{A}$ and $\alpha_{B}$ are RA values, and 
the $\delta_{A}$ and $\delta_{B}$ are Dec values of neighbouring cores. 
The last equation is the cosine rule, with $a$ and $b$ being the distances from the Sun to adjacent cores (140 pc) and $c$ being the distance between the cores.\\

\begin{figure*}[b]
    \centering
    \includegraphics[width=11 cm, height=10 cm]{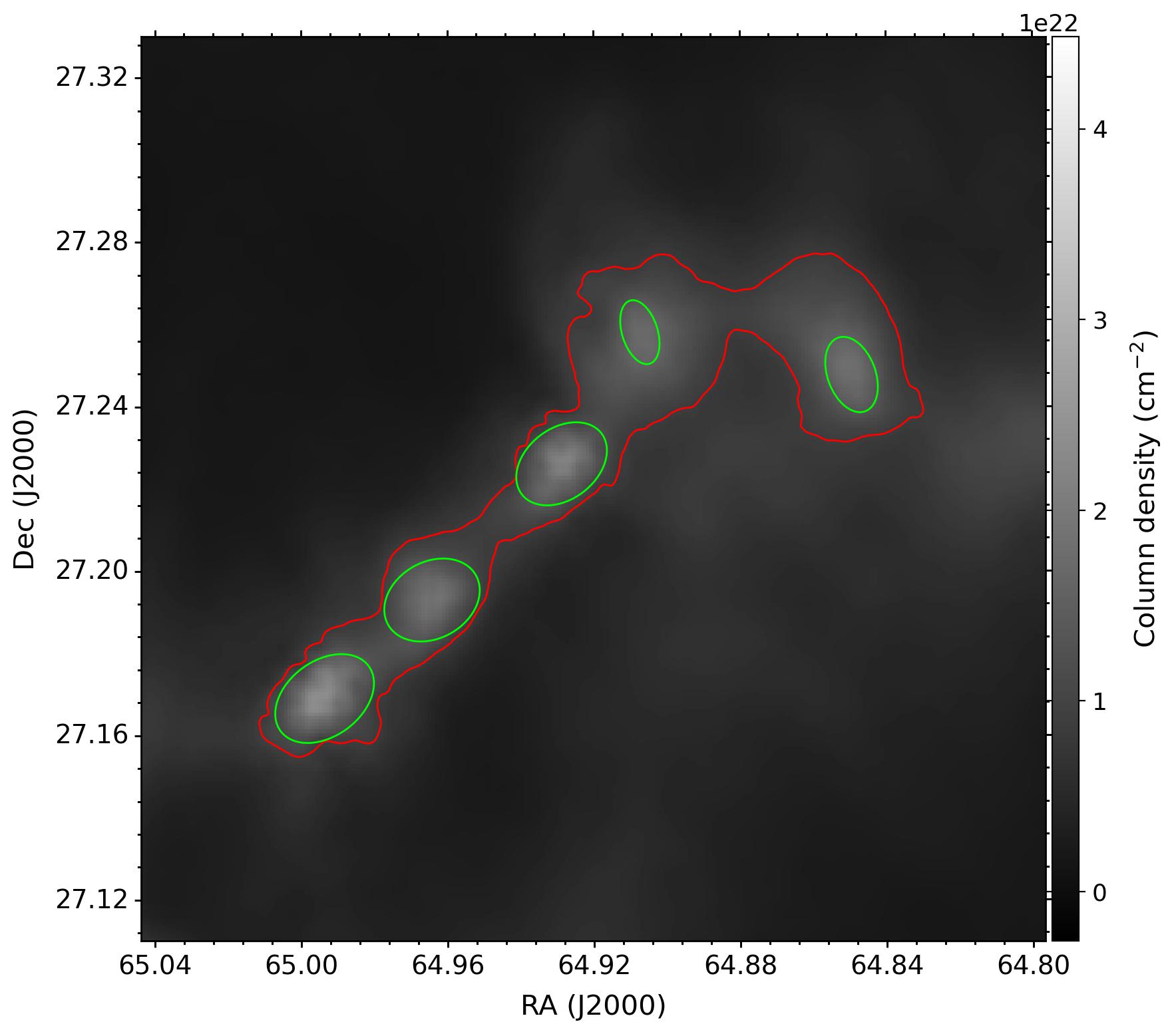}
    \caption{Herschel column density map of Taurus B213 region. The red contour is the Herschel $\mathrm{8.55\times10^{21}}$ $\mathrm{cm^{-2}}$ column density contour. The green contours represent the core boundaries based on ellipse-fitting to the cores in the POL-2 Stokes I-map.}
    \label{fig:ellipses}
\end{figure*}

\begin{figure*}[b]
    \centering
    \includegraphics[width=10 cm, height=10 cm]{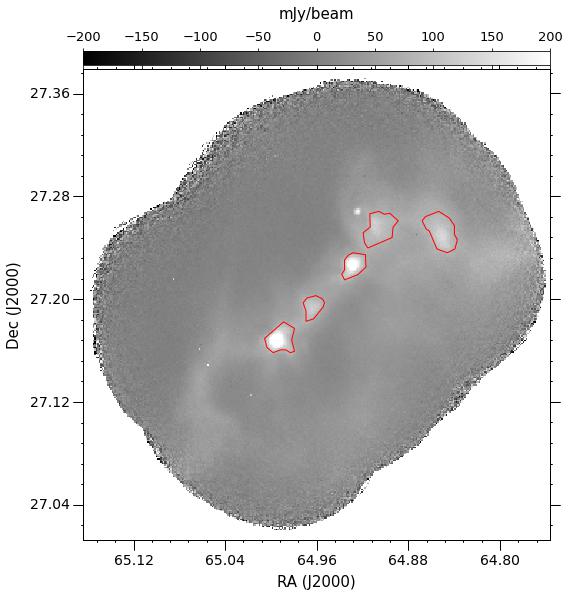}
    \caption{J-comb flux density map (obtained by combining JCMT, {\it Herschel}, {\it Planck} maps)- The five cores are shown as red polygons obtained by running the FellWalker algorithm again.}

\label{fig:combined}
\end{figure*}

These equations yield the core separations as 0.098, 0.111, 0.089, and 0.129 pc (in the order from K04169 to HGBS-1;  see Figure \ref{fig:polygon}). The total length of the filament is thus $\sim$0.42 pc. The Herschel column density contour of $8.55\times10^{21}$ $\mathrm{cm^{-2}}$ (corresponds to $\sim$5$\sigma$ of the nearby background) encompasses the five cores and the B213 filament well (see Figure \ref{fig:ellipses}). We can then obtain the filament's mass from the J-comb combined map by using equation \ref{eq:fluxmass}.\\

The integrated flux across the filament obtained from the J-comb map is 18876 mJy. Thus, the filament's mass is 7.0$\pm$3.6 $\mathrm{M_{\odot}}$. The line mass of the filament, i.e., the ratio of the total mass (7 $\mathrm{M_{\odot}}$) to its length (0.42 pc), is $\sim16.8\pm8.6$ $\mathrm{M_{\odot}}$ pc$^{-1}$. The critical mass of the filament is estimated using the following equation

\begin{equation}
    \mathrm{\left(M/L\right)_{crit} = \frac{2c_{s,fil}^2}{{G}}}.
    \label{eq:ostriker}
\end{equation}
If the filament's mass exceeds the critical mass, it undergoes fragmentation \citep{Ostriker1964}. The critical mass is estimated to be $\sim$16.1$\pm$0.6 $\mathrm{M_{\odot}}$ pc$^{-1}$, which is slightly smaller than the filament's line mass, and thus the B213 filament can fragment into cores.\\

We further examine whether the B213 filament has undergone thermal Jeans fragmentation. To estimate the Jeans length and Jeans mass, we calculate the isothermal speed of sound using the following equation 
\begin{equation}
    \mathrm{c_{s,fil}=\sqrt{\frac{k_{B}T_{d,fil}}{\mu m_{H}}}}. 
    \label{eq:csfil}
\end{equation}
Here, $\mathrm{T_{d,fil}}$ is the mean dust temperature of
11.76$\pm$0.43 K, which is the arithmetic mean of dust temperature values of pixels within the Herschel column density contour of $\mathrm{8.55\times10^{21}}$ $\mathrm{cm^{-2}}$. This $\mathrm{T_{d,fil}}$ yields an isothermal speed of 0.186$\pm$0.003 km/s.\\

The filament can be modeled as an isothermal cylinder, with its length equal to the height of the cylinder and the effective radii as the radius of the cylinder. This yields a radius of 0.025 pc (a diameter of 0.05 pc is obtained by averaging the width of the filament calculated using line segments perpendicular to the filament long axis and along the center of the cores) and a height of 0.42 pc. We then calculate the mass density using the following relation:

\begin{equation}
    \rho_{fil} = \frac{M_{fil}}{\pi R_{fil}^2H_{fil}},
    \label{eq:massdens}
\end{equation}

where $\mathrm{R_{fil}}$ is the radius of the filament, and $\mathrm{H_{fil}}$ is the height of the filament. Using this, we estimate the mass density of the filament to be $5.529\times10^{-16}$ kg $\text{m}^{-3}$. We then calculate the Jeans mass and Jeans length using the following relations:

\begin{subequations}
    \begin{equation}
         \lambda_J = c_{s,fil}\left(\frac{\pi}{G\rho_{fil}}\right)^{0.5}
    \end{equation}
    and 
    \begin{equation}
         M_J = \frac{4}{3}\pi\rho_{fil}\left(\frac{\lambda_J}{2}\right)^3, 
    \end{equation}
    \label{eq:jeans}
\end{subequations}

where G is the gravitational constant. These relations yield a Jeans length of 0.05$\pm$0.01 pc and a Jeans mass\footnote{The uncertainties in the Jeans properties are estimated by propagating the error from the filament mass and the dust temperature.} of $0.7\pm0.2\,\text{M}_{\odot}$.\\

A sophisticated magnetic version of the Jeans length and Jeans mass are given to be \citep{Krumholz&Federrath2019} 
\begin{equation}
\lambda_{\mathrm{J,\,mag}} = \left[\frac{\pi \,{c_{s}}^{2}\,(1+\beta^{-1})}{G\,\rho}\right]^{\frac{1}{2}} = \lambda_{J} (1 + \beta^{-1})^{\frac{1}{2}}
\end{equation} 
and 
\begin{equation}
M_{\mathrm{J,\,mag}} = M_{J} (1 + \beta^{-1})^{\frac{3}{2}},
\end{equation}
respectively, where $\mathrm{\lambda_{J}}$ and $\mathrm{M_{J}}$ are the standard Jeans length and Jeans mass calculated purely based on the thermal contribution. In the above equations, the term ($1+\beta^{-1}$) signifies the combined contributions of thermal and magnetic pressures.\\

%
%
%
%
%

For an isothermal equation of state, the plasma $\beta$, i.e., the ratio of thermal to magnetic pressure is 2$\mathcal{M}_{A}^{2}$/$\mathcal{M}^{2}$ \citep{Federrath&Klessen2012}. Here, the Alfv\'{e}n Mach number $\mathcal{M}_{\mathrm{A}}$ = $\mathrm{\sigma_{NT}/V_{A}}$
and the sonic Mach number $\mathcal{M} = \mathrm{\sigma_{NT}/c_{s}}$. $\mathrm{V_{A}}$ is the Alfv\'{e}n velocity which is equal to $\mathrm{\frac{B}{\sqrt{4\pi\rho}}}$, where B is the magnetic field. $\mathrm{\sigma_{NT}}$ is the non-thermal velocity dispersion, which can be computed by subtracting the thermal contribution using $\mathrm{\sigma_{NT}\,=\sqrt{\sigma_{\mathrm{tot}}^2-\sigma_{\mathrm{th}}^2}}$. The thermal velocity dispersion can be calculated by using $\mathrm{\sigma_{th}=\sqrt{\frac{k_{B}T_{d,fil}}{\mu m_{H}}}}$ \citep{Feng-Wei2023}.\\

$\mathrm{\sigma_{tot}}$ is the total velocity dispersion obtained using $\mathrm{^{13}CO}$ spectra by \citet{Chapman2011}. We consider an average magnetic field value of 40 $\mu$G in accordance with the range $\sim\,30-50\,\mu$G estimated by \citet{Chapman2011} for the B213 filament based on optical and near-infrared polarization observations. All these values along with $\lambda_{\mathrm{J,\,mag}}$ 
and $M_{\mathrm{J,\,mag}}$ are tabulated in Table \ref{tab:alfven}.

\begin{table}[h]
\centering
\caption{Computed filament quantities}
\label{tab:alfven}
\begin{tabular}{|c|c|}
\hline
\textbf{Quantity} & \textbf{Value with units} \\ \hline
$\mathrm{\sigma_{tot}}$   & 0.85$\pm$0.01 km $\mathrm{s^{-1}}$ \\ \hline
$\mathrm{\sigma_{NT}}$   & 0.83$\pm$0.01 km $\mathrm{s^{-1}}$ \\ \hline
$\mathrm{V_A}$   & 0.15 km $\mathrm{s^{-1}}$ \\ \hline
$\mathrm{\beta}$   & 3.0$\pm$0.1 \\ \hline
$\mathrm{\lambda_{J,mag}}$   & 0.06$\pm$0.02 pc \\ \hline
$\mathrm{M_{J,mag}}$   & 1.1$\pm$0.3 $\mathrm{M_{\odot}}$ \\ \hline
\end{tabular}
\end{table}

We also estimate the average mass of the five cores (obtained from Table \ref{tab:correctedmasstable}) as $\mathrm{0.8\pm0.2\,M_{\odot}}$. The average inter-core separation is $\mathrm{0.10\pm0.01\,pc}$. This distance is the value projected in 2D space. To obtain the 3D distance, we de-project it by dividing it by a factor\footnote{Following Section 5.4.1 of \citet{Sanhueza2019}, in which they state ``The average value for cos(i), with i the angle between the core separation and the observed projected separation, is given by\\
\begin{equation*}
    \hspace{3cm}\frac{1}{\pi}\int_{-\frac{\pi}{2}}^{\frac{\pi}{2}}\mathrm{cos(x)dx} = \frac{2}{\pi}\hspace{3cm}
\end{equation*}\\
On average, the observed separation will be $2/\pi$ times smaller
than the unprojected one."} of $2/\pi$. This then yields the 3D unprojected inter-core separation as $\mathrm{0.16\pm0.02\,pc}$.

\vspace{-1 em}

\section{Discussion}\label{sec:Discussion}

The calculations show that Jeans instability is an important factor in the fragmentation of this filament to protostellar cores. The calculated mean mass of the cores ($\mathrm{0.8\pm0.2\,M_{\odot}}$) closely matches with the Jeans mass ($0.7\pm0.2\,\text{M}_{\odot}$). A more detailed prescription incorporating the magnetic contributions provides a larger value of magnetic Jeans mass ($1.1\pm0.3\,\text{M}_{\odot}$). Within the uncertainty, this value matches the observed value. This modification could hint at the importance of considering magnetic field effects in the early stages of fragmentation.\\

The predicted critical wavelength $\mathrm{\lambda_{crit}}$ obtained from the relation given by \citet{Stodolkiewicz} for zero magnetic  field intensity is 1.77 times $\mathrm{\lambda_J}$. For B213, this value turns out to be 0.09$\pm$0.02 pc. A comparison of predicted and observed quantities clearly ascertains that the predicted critical length scale ($\mathrm{\lambda_{crit}}$) is smaller than the mean unprojected inter-core separation of 0.16 pc. 
This could imply that the length scale of fragmentation has increased due to the presence of strong magnetic fields perpendicular to the filament, which is in accordance with \citet{Hanawa2017}. The modified magnetic Jeans length is also smaller than the mean inter-core separation, suggesting the effect of B-fields in the later stages of fragmentation.\\

To support this scenario, we briefly state recent results of studies focussing on magnetic fields in this filament. Based on optical and near-infrared polarimetric observations, \citet{Chapman2011} estimated the plane-of-sky component of the magnetic field strength in the B213 filament to be $\sim$30~--~50 $\mu$G. They found that the relatively low-density regions are magnetically supported. These strong fields could be important in forming the B213 filament and, subsequently, its fragmentation into several cores. Furthermore, the offset angles between the filament and the large-scale magnetic field and between the cores and the large-scale magnetic field are found to be either parallel or perpendicular (cf. Table \ref{tab:coreparameters}), thereby attesting to the importance of magnetic fields in the formation of the filament and cores \citep[e.g.,][]{Li2013,Zhang2014}. Similar trends have been witnessed but at large scales such that low-density striations tend to be parallel to the magnetic fields, whereas the high-density filaments are oriented perpendicular to the magnetic fields \citep{Soler2016,Planck2016a,Wang2019}.\\

Additionally, \citet{Eswaraiah} estimated the mass-to-magnetic flux ratios, magnetic energies, and rotational energies of three cores in the B213 filament. The mass-to-flux ratios ($\lambda$) were found to be $\sim$1 for cores K04166, K04199, and $\sim$3 for Miz-8b. They showed that the three cores exhibit diverse magnetic field properties and concluded that B fields could marginally support K04166 and K04199. They also reported that the rotational energies are much smaller than the magnetic energies ($\mathrm{E_{rot}/E_{mag}<<1}$) and are thus too weak to modify the orientation of the core-scale B-fields.\\



We also find that the fragmented cores in B213 exhibit an aligned distribution with respect to their parental filament and the large-scale magnetic field. 
In related work, \citet{Tang2019} found that ordered and dynamically important magnetic fields produced well-organized and aligned dense cores in the MM2 clump of the massive Infrared Dark Cloud (IRDC) G34.43+00.24. Numerical work using magneto-hydrodynamic simulations \citep{Fontani2016} has also shown similarly aligned clump fragments in filamentary fashion in a cloud permeated by strong and ordered magnetic field lines. If turbulence dominated the magnetic fields, we could have observed a clustered fragmentation \citep[e.g,][]{Fontani2016,Tang2019}. Therefore, turbulence may not be more important than the magnetic fields in the fragmentation of the B213 filament.\\

We thus propose that the fragmentation of the B213 filament may have occurred due to thermal Jeans instability and sinusoidal perturbations in the presence of strong large-scale magnetic fields perpendicular to the filament. Such perturbations may have occurred due to 
mass accretion onto the B213 filament along the magnetic field lines \citep{Shimajiri2019}. \citet{Chapman2011} noted that an expanding bubble of a supernova remnant could have caused the arc-like shape seen northwest of the filament \citep{Shimajiri2019}. The detection of the gamma-ray pulsar FermiLAT PSR J0357+32 \citep{Abdo2009} supports this claim. This supernova remnant might have also contributed to perturbing  the equilibrium of the B213 filament.



\vspace{-1em}
\section{Summary}\label{sec:Summary}

\begin{itemize}
    \item The B213 filament is collapsing into Sun-type star-forming cores. The calculated Jeans mass closely matches the mean mass of the cores.
    \item The mean inter-core separation is larger than the critical length scale. This may be due to the presence of strong magnetic fields perpendicular to the filament that could cause the fragmentation length scale to increase. This hints at a thermal Jeans-instability and sinusoidal perturbation-driven fragmentation mode in the presence of dynamically important magnetic fields.
    \item Although this scenario can be adopted as a useful model, deviations may be attributed to core-scale magnetic fields, inhomogeneous accretion, and turbulence.
\end{itemize}


\vspace{-2em}
\section*{Acknowledgements}
The authors thank the anonymous reviewer for their detailed and insightful comments, which have improved the overall flow and contents of the manuscript. RA carried out this work for the Monsoon 2021 semester project under the supervision of CE at the Indian Institute of Science Education and Research (IISER), Tirupati. CE acknowledges the financial support from grant RJF/2020/000071 as a part of the Ramanujan Fellowship awarded by the Science and Engineering Research Board (SERB).
\vspace{-1em}

\balance



\bibliography{main.bbl}

\end{document}